\documentclass[english,prb,reprint,a4paper,showpacs]{revtex4-1}
\pdfoutput=1
\usepackage[scaled=0.92]{helvet}
\usepackage{courier}
\usepackage[T1]{fontenc}
\usepackage[latin9]{inputenc}
\usepackage[pdftex]{color}
\usepackage{babel}

\usepackage[pdftex]{graphicx}
\usepackage[unicode=true, pdfusetitle,
 bookmarks=true,bookmarksnumbered=false,bookmarksopen=false,
 breaklinks=false,pdfborder={0 0 0},backref=false,colorlinks=true]
 {hyperref}

\makeatletter

\providecommand{\tabularnewline}{\\}
\newcommand{\lyxdot}{.}

\@ifundefined{textcolor}{}
{%
 \definecolor{BLACK}{gray}{0}
 \definecolor{WHITE}{gray}{1}
 \definecolor{RED}{rgb}{1,0,0}
 \definecolor{GREEN}{rgb}{0,1,0}
 \definecolor{BLUE}{rgb}{0,0,1}
 \definecolor{CYAN}{cmyk}{1,0,0,0}
 \definecolor{MAGENTA}{cmyk}{0,1,0,0}
 \definecolor{YELLOW}{cmyk}{0,0,1,0}
 }

\usepackage{fouriernc}

\makeatother

\begin{document}

\title{First-principles study for the adsorption of segments of BPA-PC on
$\alpha$-Al$_{2}$O$_{3}$(0001)}

\author{Janne Blomqvist and Petri Salo}

\affiliation{Department of Applied Physics, School of Science, Aalto University,
P.O. Box 11100, FI-00076 Aalto, Finland}
\begin{abstract}
We have studied the adsorption of bisphenol-A-polycarbonate (BPA-PC)
on the $\alpha$-Al$_{2}$O$_{3}$(0001) surface using density-functional
theory (DFT) with van der Waals (vdW) corrections. The BPA-PC polymer
can be divided into its chemical fragments which are phenylene, carbonate
and isopropylidene groups. We have calculated the adsorption energy
and geometry of the BPA-PC segments that consist of two to three adjacent
groups of the polymer. Our DFT results show that the adsorption is
dominated by the vdW interaction. It is also important to include
the interaction of nearest-neighbor groups in order to provide a realistic
environment for the adsorption of the polymer onto the surface. Our
results also show that the BPA-PC molecule attaches to the alumina
surface via the carbonate group located in the middle of the molecule
chain.
\end{abstract}


\maketitle
The bisphenol-A-polycarbonate (BPA-PC) molecule studied in this work
is an important industrial polymer used in different composite materials
together with, e.g., metals. The molecule has extensively been studied
both experimentally \cite{jho1991,kim1992,floudas1993,katana1993,buchenau1994,eilhard1999}
and computationally \cite{eilhard1999,abrams2003a,abrams2003b,akola2002,akola2003,delle_site2002}.
The interaction of BPA-PC or its chemical groups, phenylene, carbonate,
and isopropylidene, with crystal surfaces has also been studied, and
they provide a good model system for studying the properties of surfaces
and interfaces \cite{abrams2003b,blomqvist2009,chakarova2006,delle_site2002,delle_site2003,delle_site2004,delle_site2005,johnston2007,johnston2011}.

Density-functional theory (DFT) has become an important tool for the
study of matter at the electronic-structure level, including processes
on surfaces such as the adsorption of atoms or molecules. Even though
the covalent interaction is crucial for the intramolecular binding
or the binding of single atoms on surfaces, large molecules with closed
electron shells are interacting relatively weakly with the surfaces.
Large, especially organic, molecules interact with the surface mostly
via the van der Waals (vdW) interaction. This electric dipole-dipole
interaction has a longer range compared to the covalent interaction,
allowing the molecule to interact with the surface comparatively far
from it \cite{moses2009,wellendorff2010}. However, the conventional
DFT includes only the covalent interactions and thus it fails to give
the correct adsorption energy of large molecules on surfaces. Recent
developments have made it possible to include the effects of vdW interactions
in the DFT description of the large molecule systems. 

In this work we extend a subset of our previous DFT calculations for
the groups of BPA-PC (Ref. \cite{blomqvist2009}) by taking into account
the interactions of nearest-neighbor groups as well as the vdW interactions.
Both of these factors can be crucial in understanding the adsorption
processes of large molecules, with macromolecules (polymers) being
an extreme example. To show this, we have calculated the adsorption
energy and geometry of different segments consisting of two to three
adjacent groups of the BPA-PC polymer chain on the $\alpha$-alumina
(0001) surface. 

In the numerical calculations we combine two different DFT approaches
depending on the purpose described below. Compared to adsorption calculations
for small molecules or individual atoms, adsorption calculations for
large molecules offer a number of challenges. First, in order to fit
the molecule into the simulation box such that it does not interact
with its own mirror image due to periodic boundary conditions, the
box has to be large. Second, large molecules, especially segments
of polymers studied here, have large conformational degrees of freedom,
meaning that relaxing them requires a lot of time. These two issues,
combined with the fact that one needs a similar level of accuracy
as for normal adsorption calculations, essentially makes a traditional
plane-wave DFT approach unfeasible. In order to perform these calculations,
we have to make quick scans with a relative fast linear combination
of atomic orbitals (LCAO) DFT method first and, after that, refining
the interesting cases with a more accurate but significantly more
expensive real-space finite-difference DFT method. This combining
of two codes is significantly easier by controlling both of them via
a unified high-level interface \cite{bahn2002}.

For the LCAO calculations, we used the Siesta code \cite{soler2002},
using single-$\zeta$ polarized (SZP) orbitals and a single k-point
at the $\Gamma$ point. The many-body effects were approximated with
the revised version \cite{hammer1999} of the Perdew-Burke-Ernzerhof
\cite{perdew1996,perdew1996a} (RPBE) form of the generalized-gradient
approximation (GGA). The atomic coordinates were relaxed until the
maximum force was less than $0.04$ eV/Å. The finite-difference calculations
were done using projector augmented waves (PAW) \cite{blochl1994}
as implemented in the real-space finite-difference code GPAW \cite{mortensen2005,bahn2002,enkovaara2010}.
We used a grid spacing of approximately $0.19$ Å, and a 2x1x1 k-point
mesh using the Monkhorst-Pack scheme \cite{monkhorst1976,pack1977}.
The exchange-correlation and vdW interactions were taken into account
with a revised version of the PBE GGA \cite{perdew1996,perdew1996a,zhang1997}
(revPBE), and a self-consistent implementation of a fast Fourier transform
(FFT) based method \cite{roman-perez2009} based on the vdW-functional
of Dion et al. \cite{dion2004,dion2005}. The systems were relaxed
until the maximum force on all atoms were less than $0.02$ eV/Å.
The Fermi level was smeared using Fermi smearing with a smearing parameter
of 0.1 eV.

The $\alpha$-Al$_{2}$O$_{3}$(0001) supercell contained 96 Al and
144 O atoms in 12 Al and 6 O layers forming an 8x4 hexagonal surface
supercell with Cartesian dimensions 8.3 Å x 19.1 Å x 27.0 Å. The height
of the slab was 12.1 Å, and there was a 14.9 Å vacuum region. The
lowest Al and O layers were fixed, while the rest of the slab was
relaxed. The surface was Al-terminated, as previous calculations have
shown this to be the most stable termination \cite{guo1992}, although
there are some conflicting experimental results suggesting a preference
for a mixture of different terminations \cite{toofan1998}. For some
calculations, depending on the adsorbate configuration on the surface,
the cell size was doubled to 16.6 Å x 19.1 Å x 27.0 Å.

\begin{figure}
\begin{centering}
\includegraphics{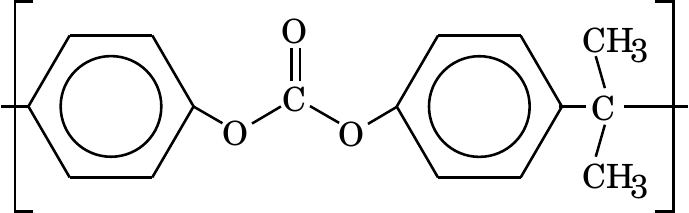}
\par\end{centering}

\caption{\label{fig:BPA-PC-monomer}Structural formula of the BPA-PC monomer.
The monomer consists of the groups P, C, P, and I, respectively. }

\end{figure}

A BPA-PC monomer can be seen in Fig. \ref{fig:BPA-PC-monomer}. The
monomer consists of three different chemical groups, phenylene (marked
with P, chemical formula C$_{6}$H$_{4}$), carbonate (C, CO$_{3}$),
and isopropylidene (I, C$_{3}$H$_{6}$). As the BPA-PC polymer chains
are usually terminated at both ends by a P group \cite{delle_site2002},
for the adsorption a PC segment was simulated. The adsorption of the
PI segment has not been simulated, as the I group sterically prevents
the P group from getting close to the surface. An IC segment has not
been studied either, because it does not contain any P group. For
the intra-chain adsorptions, segments consisting of three groups were
simulated, but the chains of length of four groups or longer have
not been simulated because of too time-consuming calculations. Thus,
the segments studied are PC, PCP, IPC, and PIP. The dangling bonds
of the neighbor groups have been passivated with H atoms. If single
groups are simulated, the groups will be benzene (C$_{6}$H$_{6}$),
carbonic acid (H$_{2}$CO$_{3}$), and propane (C$_{3}$H$_{8}$)
due to the H passivation. 

\begin{figure}
\begin{centering}
\includegraphics{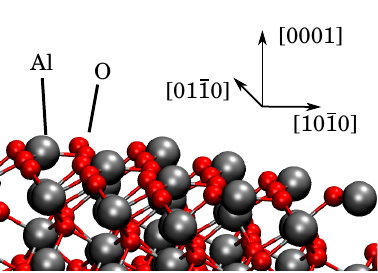}~\includegraphics{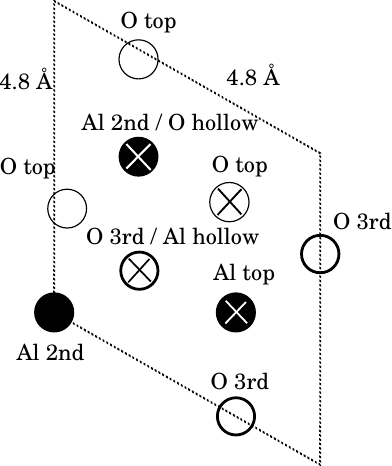}
\par\end{centering}

\caption{\label{fig:Adsorption-sites-Al2O3}General view (left) and surface
atoms (right) on the Al-terminated $\alpha$-Al$_{2}$O$_{3}$(0001)
surface. On the left, Al atoms are larger gray spheres and O atoms
smaller red spheres (dark gray spheres in b/w). On the right, Al atoms
are filled circles and O atoms empty circles. Due to the Al$_{2}$O$_{3}$
surface relaxation the uppermost Al and O layer are almost at the
same height, then the second layer with Al atoms, and finally the
third layer with O atoms. A 2x2 surface supercell box can also be
seen.}

\end{figure}

First the minimum-energy adsorption configurations for each segment
PC, PCP, IPC, and PIP on the $\alpha$-Al$_{2}$O$_{3}$(0001) surface
were searched for. The different adsorption sites used, the Al top,
Al hollow, O top and O hollow sites, can be seen in Fig. \ref{fig:Adsorption-sites-Al2O3}.
The adsorption site for a large molecule was determined so that we
picked from the molecule an atom which we located at the adsorption
site in each calculation. For PC and IPC the atom at the adsorption
site was one of the C atoms in the P group and for PCP the atom was
the O atom of the C group closest to the surface. For PIP there were
two possible configurations, either one of the C or H atoms in the
I group has been placed at the adsorption site. Furthermore, the rotation
of the segment was chosen keeping in mind the steric constraints that
a BPA-PC polymer chain could have had. Each segment was rotated at
30$^{\circ}$ intervals, resulting in a total of 240 configurations.
Symmetrically non-equivalent configurations were evaluated with the
Siesta code using the fast LCAO method. 

The Siesta results are quite qualitative and the differences between
different configurations of the same system are some tenths of an
electronvolt. This is to be expected since the large size of the segments
means that they will not fit into a hollow site like an atom or a
smaller molecule can do, and it also means that the notion of an adsorption
site is at best approximate as the molecule segment interacts with
the surface over a large area. The minimum adsorption configuration
of the segment for each adsorption site is the O hollow site for PC
and PCP, and the O top site for IPC and PIP with the C atom at the
adsorption site. These systems have been studied further with more
accurate, but time-consuming, finite-difference PAW scheme, including
vdW interactions. 

\begin{figure}
\begin{centering}
\includegraphics[width=0.35\columnwidth]{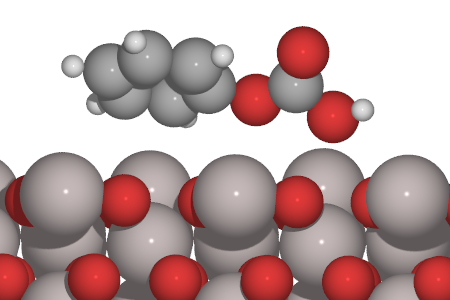}~\includegraphics[width=0.35\columnwidth]{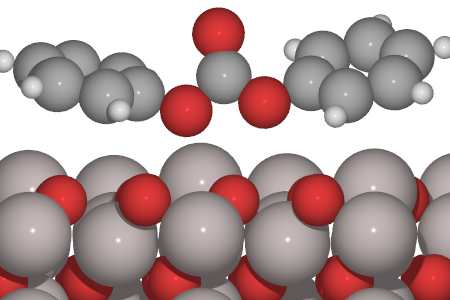}
\par\end{centering}

\begin{centering}
\includegraphics[width=0.35\columnwidth]{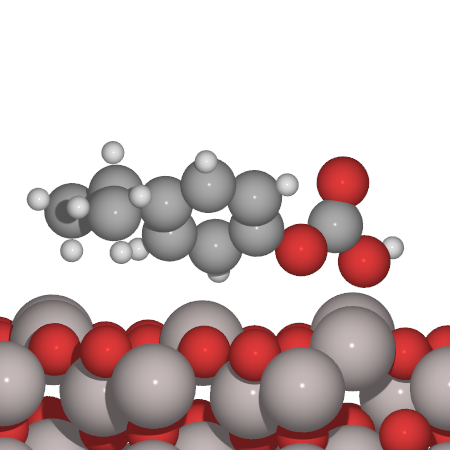}~\includegraphics[width=0.35\columnwidth]{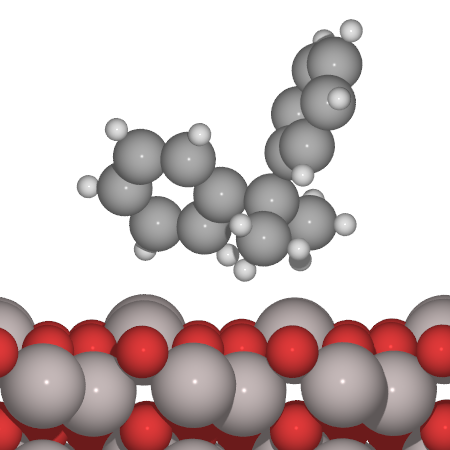}
\par\end{centering}

\caption{\label{fig:Equilibrium-adsorption-configurations}Equilibrium adsorption
configurations for the segments on the surface. The segments are PC
(top left), PCP (top right), IPC (bottom left), and PIP (bottom right).
Red spheres (dark gray spheres in b/w) represent O atoms, large gray
spheres Al atoms, gray spheres C atoms, and small gray spheres represent
H atoms.}

\end{figure}

The results of the GPAW calculations using the revPBE-vdW functional
are shown in Table \ref{tab:vdw-gpaw-adsorption-energies}. One can
see that the vdW contribution is very important, and especially for
the adsorption of PIP on the alumina surface there is, in practice,
no covalent interaction at all. The adsorption energies excluding
vdW interactions are, at most, around half of the vdW contribution,
confirming the hypothesis that including vdW interactions is essential
for calculating large molecules on surfaces. In Fig. \ref{fig:Equilibrium-adsorption-configurations}
one can see the adsorption geometries for the different segments on
the surface.

\begin{table}
\caption{\label{tab:vdw-gpaw-adsorption-energies}Calculated adsorption energies
with PAW vdW. The first column refers to the adsorption energy using
self-consistent vdW DFT, the second column is the adsorption energy
without the vdW contribution, and the third column is the vdW contribution
to the adsorption energy.}

\centering{}\begin{tabular}{c|ccc}
\hline 
 & $E_{\mathrm{ads}}$(eV) & $E{}_{\mathrm{ads,revPBE}}$ (eV) & $E{}_{\mathrm{vdw}}$ (eV)\tabularnewline
\hline
PC & $1.22$ & $0.42$ & $0.80$\tabularnewline
PCP & $1.49$ & $0.22$ & $1.26$\tabularnewline
IPC & $1.47$ & $0.32$ & $1.15$\tabularnewline
PIP & $0.90$ & $0.03$ & $0.88$\tabularnewline
\hline
\end{tabular}
\end{table}

One can see that the segments that contain the C group have larger
adsorption energies (see Table \ref{tab:vdw-gpaw-adsorption-energies}).
The C group binds to the surface more eagerly than the other parts
that can also be seen in Fig. \ref{fig:Equilibrium-adsorption-configurations}.
The surface Al atom underneath the O atom in the C group in each case
has been pulled upward by about 0.2 Å compared to the clean surface.
This result is also consistent with the results in Refs. \cite{blomqvist2009,chakarova2006}
where it was found that phenol (C$_{6}$H$_{5}$OH) and carbonic acid
binds to the Al$_{2}$O$_{3}$ surface via the oxygen atom in the
phenol group and one of the oxygen atoms in the carbonic-acid group.

The neighboring groups are also important. The adsorption energy of
the PC segment increases when the molecule chain contains an extra
P forming a PCP segment on the surface. However, the increase is almost
the same if one looks at the change of the adsorption energy between
PC and IPC segments. As the PCP segment, the IPC segment interacts
with the alumina surface via the C group. The P group seems to be
inert to the alumina surface, which is consistent with the results
in Ref. \cite{blomqvist2009} where it was found that the benzene
group does not bind to the alumina surface. In this work, the PIP
segment that does not contain the C group interacts with the alumina
surface via the I group and the P groups are only bound to the I group
in the middle (see Fig. \ref{fig:Equilibrium-adsorption-configurations}).
However, if the chain contains four groups or more, it will contain
at least one C group that will attach to the alumina surface. Unfortunately,
the DFT calculations for chains with four groups or more become too
time-consuming to perform. As the BPA-PC polymer chains are usually
terminated at both ends by a P group \cite{delle_site2002}, we can
conclude that this polymer chain attaches to the alumina surface via
the C group in the middle of the polymer. 

\begin{figure}
\begin{centering}
\includegraphics{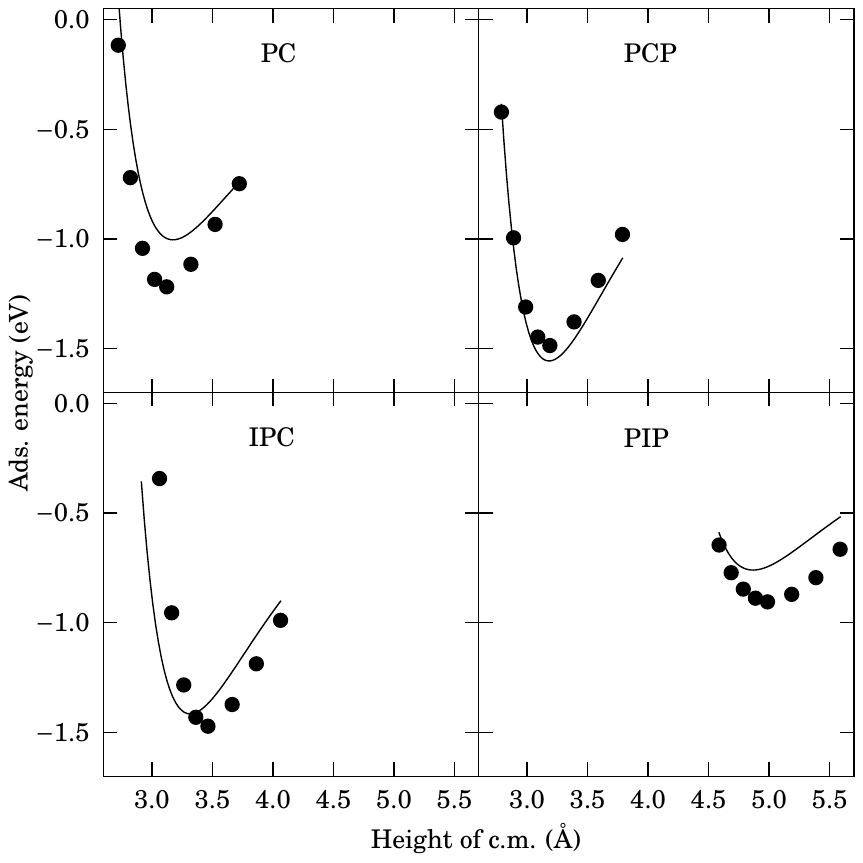}
\par\end{centering}

\caption{\label{fig:ads-en-fz}Adsorption energies as a function of the height
of the segment center-of-masses from the surface. The solid line for
each case represents an LJ fitting to the data.}

\end{figure}

The adsorption energy of these segments as a function of the perpendicular
distance from the surface has also been calculated. For these calculations
no geometry optimization was done, only the \textit{z} coordinate
of all the atoms in the segment was changed. These results can be
seen in Fig. \ref{fig:ads-en-fz}, in which the adsorption energies
of the segments are presented as a function of the height of the center
of mass (c.m.) of the segment from the surface atomic layer. One can
see that the equilibrium distance of the segment from the surface
is pretty much the same, except for the PIP segment. This partly explains
the smaller adsorption energy of PIP on the alumina surface.

We have also made a fit to the data in Fig. \ref{fig:ads-en-fz} using
a 10-4 Lennard-Jones (LJ) potential\[
E_{\mathrm{LJ}}^{10-4}(z)=\frac{5}{3}\epsilon\left[\frac{2}{5}\left(\frac{\sigma}{z}\right)^{10}-\left(\frac{\sigma}{z}\right)^{4}\right]\,,\]
which represents a 12-6 LJ potential integrated over the first atomic
plane in the surface. In the fitting procedure, we have fitted sums
of the single group potentials, that is, P+C, P+C+P, I+P+C, P+I+P,
and minimized, in the least-squares sense, the difference between
these sums and the vdW DFT segment data by varying the $\epsilon$
and $\sigma$ parameters for P, C, and I. The LJ fitting parameters
are shown in Table \ref{tab:lennard-jones-parameters}. For comparison,
we have also listed in Table \ref{tab:lennard-jones-parameters} the
average adsorption energies and heights of single groups on the surface.
Single groups were simulated in a fixed conformation as they were
in the segments, and each of the dangling bonds of the groups was
passivated with hydrogen atoms. These results also show that the adhesion
of the C group to the alumina surface is the strongest. When one compares
the LJ fitting parameter and the single-group results, the well depth
$\epsilon$ of the LJ potential for the I group is not in line with
the other results. This discrepancy can be understood if one looks
at the PIP segment on the surfaces. The adhesion of PIP on the surface
is mostly due to the I group and that is why the adsorption energy
of a single I group in the fitting is so high. 

The LJ fitting has also been shown in Fig. \ref{fig:ads-en-fz}. The
fitting seems reasonably good, considering the simplicity of the potential
function, and the fitting is the same for all the cases. In particular,
the fitting is good for the IPC and PCP segments, which are the most
important cases, as PC is only at the chain end, and PIP is adsorbed
far away from the surface. Since the LJ potential is a good approximation
for the vdW interactions, the fitting confirms that the interaction
between these segments and the alumina surface is mainly due to the
vdW interactions, in agreement with Table \ref{tab:vdw-gpaw-adsorption-energies}. 

\begin{table}
\caption{\label{tab:lennard-jones-parameters}Fitted LJ parameters and single
group averaged vdW DFT results.}

\centering{}\begin{tabular}{c|cccc}
\hline 
 & $\epsilon$ (eV) & $\sigma$ (Å) & $E{}_{\mathrm{ads}}$(eV) & $z$ (Å)\tabularnewline
\hline
P & $0.50$ & $3.02$ & $0.30$ & $3.54$\tabularnewline
C & $0.57$ & $3.12$ & $0.59$ & $3.05$\tabularnewline
I & $0.50$ & $3.43$ & $0.18$ & $3.59$\tabularnewline
\hline
\end{tabular}
\end{table}

As a summary, we have calculated multi-group segments of the BPA-PC
on an aluminum oxide surface using vdW DFT. The results show that,
for these large segments, the vdW interaction plays a major role in
the adhesion process at the atomic scale. Similarly, in order to provide
a more realistic environment for each group on the surface, we have
to include the nearest-neighbor groups, too. This is partly due to
the long-range effect of the vdW interaction. We have also found that
the BPA-PC molecule attaches to the alumina surface via the carbonate
group located in the middle of the polymer chain. The LJ fittings
of the interaction between the polymer fragments and the alumina surface
can be used in molecular-dynamics simulations in the future to study
the properties of the interface between these materials.

This work has been supported by the Finnish Funding Agency for Technology
and Innovation (Tekes) through the K3MAT project, and by the Academy
of Finland through its Computational Nanoscience (COMP) Center of
Excellence program headed by R. Nie\-minen. We also acknowledge the
computer resources provided by the Finnish IT Center for Science (CSC).
Finally, we wish to thank K. Johnston and A. Gulans for discussions.

\bibliography{bpapc}

\begin{thebibliography}{10}%
\makeatletter
\providecommand \@ifxundefined [1]{%
 \ifx #1\undefined \expandafter \@firstoftwo
 \else \expandafter \@secondoftwo
\fi
}%
\providecommand \@ifnum [1]{%
 \ifnum #1\expandafter \@firstoftwo
 \else \expandafter \@secondoftwo
\fi
}%
\providecommand \enquote [1]{``#1''}%
\providecommand \bibnamefont  [1]{#1}%
\providecommand \bibfnamefont [1]{#1}%
\providecommand \citenamefont [1]{#1}%
\providecommand\href[0]{\@sanitize\@href}%
\providecommand\@href[1]{\endgroup\@@startlink{#1}\endgroup\@@href}%
\providecommand\@@href[1]{#1\@@endlink}%
\providecommand \@sanitize [0]{\begingroup\catcode`\&12\catcode`\#12\relax}%
\@ifxundefined \pdfoutput {\@firstoftwo}{%
 \@ifnum{\z@=\pdfoutput}{\@firstoftwo}{\@secondoftwo}%
}{%
 \providecommand\@@startlink[1]{\leavevmode\special{html:<a href="#1">}}%
 \providecommand\@@endlink[0]{\special{html:</a>}}%
}{%
 \providecommand\@@startlink[1]{%
  \leavevmode
  \pdfstartlink
   attr{/Border[0 0 1 ]/H/I/C[0 1 1]}%
   user{/Subtype/Link/A<</Type/Action/S/URI/URI(#1)>>}%
  \relax
 }%
 \providecommand\@@endlink[0]{\pdfendlink}%
}%
\providecommand \url  [0]{\begingroup\@sanitize \@url }%
\providecommand \@url [1]{\endgroup\@href {#1}{\urlprefix}}%
\providecommand \urlprefix [0]{URL }%
\providecommand \Eprint[0]{\href }%
\@ifxundefined \urlstyle {%
  \providecommand \doi [1]{doi:\discretionary{}{}{}#1}%
}{%
  \providecommand \doi [0]{doi:\discretionary{}{}{}\begingroup
  \urlstyle{rm}\Url }%
}%
\providecommand \doibase [0]{http://dx.doi.org/}%
\providecommand \Doi[1]{\href{\doibase#1}}%
\providecommand \bibAnnote [3]{%
  \BibitemShut{#1}%
  \begin{quotation}\noindent
    \textsc{Key:}\ #2\\\textsc{Annotation:}\ #3%
  \end{quotation}%
}%
\providecommand \bibAnnoteFile [2]{%
  \IfFileExists{#2}{\bibAnnote {#1} {#2} {\input{#2}}}{}%
}%
\providecommand \typeout [0]{\immediate \write \m@ne }%
\providecommand \selectlanguage [0]{\@gobble}%
\providecommand \bibinfo [0]{\@secondoftwo}%
\providecommand \bibfield [0]{\@secondoftwo}%
\providecommand \translation [1]{[#1]}%
\providecommand \BibitemOpen[0]{}%
\providecommand \bibitemStop [0]{}%
\providecommand \bibitemNoStop [0]{.\EOS\space}%
\providecommand \EOS [0]{\spacefactor3000\relax}%
\providecommand \BibitemShut [1]{\csname bibitem#1\endcsname}%
\bibitem{jho1991}%
  \BibitemOpen
  \bibfield{author}{%
  \bibinfo {author} {\bibfnamefont{J.~Y.}\ \bibnamefont{Jho}}\ and\ \bibinfo
  {author} {\bibfnamefont{A.~F.}\ \bibnamefont{Yee}},\ }%
  \bibfield{journal}{%
  \Doi{10.1021/ma00008a031}{\bibinfo {journal} {Macromolecules}}\ }%
  \textbf{\bibinfo {volume} {24}},\ \bibinfo {pages} {1905} (\bibinfo {year}
  {1991})%
  \bibAnnoteFile{NoStop}{jho1991}%
\bibitem{kim1992}%
  \BibitemOpen
  \bibfield{author}{%
  \bibinfo {author} {\bibfnamefont{C.}~\bibnamefont{Kim}}\ and\ \bibinfo
  {author} {\bibfnamefont{D.}~\bibnamefont{Paul}},\ }%
  \bibfield{journal}{%
  \Doi{10.1021/ma00038a012}{\bibinfo {journal} {Macromolecules}}\ }%
  \textbf{\bibinfo {volume} {25}},\ \bibinfo {pages} {3097} (\bibinfo {year}
  {1992})%
  \bibAnnoteFile{NoStop}{kim1992}%
\bibitem{floudas1993}%
  \BibitemOpen
  \bibfield{author}{%
  \bibinfo {author} {\bibfnamefont{G.}~\bibnamefont{Floudas}}, \bibinfo
  {author} {\bibfnamefont{T.}~\bibnamefont{Pakula}}, \bibinfo {author}
  {\bibfnamefont{M.}~\bibnamefont{Stamm}},\ and\ \bibinfo {author}
  {\bibfnamefont{E.}~\bibnamefont{Fischer}},\ }%
  \bibfield{journal}{%
  \Doi{10.1021/ma00059a028}{\bibinfo {journal} {Macromolecules}}\ }%
  \textbf{\bibinfo {volume} {26}},\ \bibinfo {pages} {1671} (\bibinfo {year}
  {1993})%
  \bibAnnoteFile{NoStop}{floudas1993}%
\bibitem{katana1993}%
  \BibitemOpen
  \bibfield{author}{%
  \bibinfo {author} {\bibfnamefont{G.}~\bibnamefont{Katana}}, \bibinfo {author}
  {\bibfnamefont{F.}~\bibnamefont{Kremer}}, \bibinfo {author}
  {\bibfnamefont{E.~W.}\ \bibnamefont{Fischer}},\ and\ \bibinfo {author}
  {\bibfnamefont{R.}~\bibnamefont{Plaetschke}},\ }%
  \bibfield{journal}{%
  \Doi{10.1021/ma00064a013}{\bibinfo {journal} {Macromolecules}}\ }%
  \textbf{\bibinfo {volume} {26}},\ \bibinfo {pages} {3075} (\bibinfo {year}
  {1993})%
  \bibAnnoteFile{NoStop}{katana1993}%
\bibitem{buchenau1994}%
  \BibitemOpen
  \bibfield{author}{%
  \bibinfo {author} {\bibfnamefont{U.}~\bibnamefont{Buchenau}}, \bibinfo
  {author} {\bibfnamefont{C.}~\bibnamefont{Sch\"{o}nfeld}}, \bibinfo {author}
  {\bibfnamefont{D.}~\bibnamefont{Richter}}, \bibinfo {author}
  {\bibfnamefont{T.}~\bibnamefont{Kanaya}}, \bibinfo {author}
  {\bibfnamefont{K.}~\bibnamefont{Kaji}},\ and\ \bibinfo {author}
  {\bibfnamefont{R.}~\bibnamefont{Wehrmann}},\ }%
  \bibfield{journal}{%
  \Doi{10.1103/PhysRevLett.73.2344}{\bibinfo {journal} {Physical Review
  Letters}}\ }%
  \textbf{\bibinfo {volume} {73}},\ \bibinfo {pages} {2344} (\bibinfo {year}
  {1994})%
  \bibAnnoteFile{NoStop}{buchenau1994}%
\bibitem{eilhard1999}%
  \BibitemOpen
  \bibfield{author}{%
  \bibinfo {author} {\bibfnamefont{J.}~\bibnamefont{Eilhard}} \emph{et~al.},\
  }%
  \bibfield{journal}{%
  \Doi{10.1063/1.477889}{\bibinfo {journal} {The Journal of Chemical Physics}}\
  }%
  \textbf{\bibinfo {volume} {110}},\ \bibinfo {pages} {1819} (\bibinfo {year}
  {1999})%
  \bibAnnoteFile{NoStop}{eilhard1999}%
\bibitem{abrams2003a}%
  \BibitemOpen
  \bibfield{author}{%
  \bibinfo {author} {\bibfnamefont{C.~F.}\ \bibnamefont{Abrams}}\ and\ \bibinfo
  {author} {\bibfnamefont{K.}~\bibnamefont{Kremer}},\ }%
  \bibfield{journal}{%
  \Doi{10.1021/ma0213495}{\bibinfo {journal} {Macromolecules}}\ }%
  \textbf{\bibinfo {volume} {36}},\ \bibinfo {pages} {260} (\bibinfo {year}
  {2003})%
  \bibAnnoteFile{NoStop}{abrams2003a}%
\bibitem{abrams2003b}%
  \BibitemOpen
  \bibfield{author}{%
  \bibinfo {author} {\bibfnamefont{C.~F.}\ \bibnamefont{Abrams}}, \bibinfo
  {author} {\bibfnamefont{L.}~\bibnamefont{Delle~Site}},\ and\ \bibinfo
  {author} {\bibfnamefont{K.}~\bibnamefont{Kremer}},\ }%
  \bibfield{journal}{%
  \Doi{10.1103/PhysRevE.67.021807}{\bibinfo {journal} {Physical Review E}}\ }%
  \textbf{\bibinfo {volume} {67}},\ \bibinfo {pages} {021807} (\bibinfo {year}
  {2003})%
  \bibAnnoteFile{NoStop}{abrams2003b}%
\bibitem{akola2002}%
  \BibitemOpen
  \bibfield{author}{%
  \bibinfo {author} {\bibfnamefont{J.}~\bibnamefont{Akola}}, \bibinfo {author}
  {\bibfnamefont{P.}~\bibnamefont{Ballone}},\ and\ \bibinfo {author}
  {\bibfnamefont{R.~O.}\ \bibnamefont{Jones}},\ }%
  \bibfield{journal}{%
  \Doi{10.1021/ma011757t}{\bibinfo {journal} {Macromolecules}}\ }%
  \textbf{\bibinfo {volume} {35}},\ \bibinfo {pages} {2327} (\bibinfo {year}
  {2002})%
  \bibAnnoteFile{NoStop}{akola2002}%
\bibitem{akola2003}%
  \BibitemOpen
  \bibfield{author}{%
  \bibinfo {author} {\bibfnamefont{J.}~\bibnamefont{Akola}}\ and\ \bibinfo
  {author} {\bibfnamefont{R.~O.}\ \bibnamefont{Jones}},\ }%
  \bibfield{journal}{%
  \Doi{10.1021/ma021630j}{\bibinfo {journal} {Macromolecules}}\ }%
  \textbf{\bibinfo {volume} {36}},\ \bibinfo {pages} {1355} (\bibinfo {year}
  {2003})%
  \bibAnnoteFile{NoStop}{akola2003}%
\bibitem{delle_site2002}%
  \BibitemOpen
  \bibfield{author}{%
  \bibinfo {author} {\bibfnamefont{L.}~\bibnamefont{Delle~Site}}, \bibinfo
  {author} {\bibfnamefont{C.~F.}\ \bibnamefont{Abrams}}, \bibinfo {author}
  {\bibfnamefont{A.}~\bibnamefont{Alavi}},\ and\ \bibinfo {author}
  {\bibfnamefont{K.}~\bibnamefont{Kremer}},\ }%
  \bibfield{journal}{%
  \Doi{10.1103/PhysRevLett.89.156103}{\bibinfo {journal} {Physical Review
  Letters}}\ }%
  \textbf{\bibinfo {volume} {89}},\ \bibinfo {pages} {156103} (\bibinfo {year}
  {2002})%
  \bibAnnoteFile{NoStop}{delle_site2002}%
\bibitem{blomqvist2009}%
  \BibitemOpen
  \bibfield{author}{%
  \bibinfo {author} {\bibfnamefont{J.}~\bibnamefont{Blomqvist}}\ and\ \bibinfo
  {author} {\bibfnamefont{P.}~\bibnamefont{Salo}},\ }%
  \bibfield{journal}{%
  \Doi{10.1088/0953-8984/21/22/225001}{\bibinfo {journal} {Journal of Physics:
  Condensed Matter}}\ }%
  \textbf{\bibinfo {volume} {21}},\ \bibinfo {pages} {225001} (\bibinfo {year}
  {2009})%
  \bibAnnoteFile{NoStop}{blomqvist2009}%
\bibitem{chakarova2006}%
  \BibitemOpen
  \bibfield{author}{%
  \bibinfo {author} {\bibfnamefont{S.~D.}\ \bibnamefont{Chakarova-Käck}},
  \bibinfo {author} {\bibnamefont{Øyvind Borck}}, \bibinfo {author}
  {\bibfnamefont{E.}~\bibnamefont{Schröder}},\ and\ \bibinfo {author}
  {\bibfnamefont{B.~I.}\ \bibnamefont{Lundqvist}},\ }%
  \bibfield{journal}{%
  \Doi{10.1103/PhysRevB.74.155402}{\bibinfo {journal} {Physical Review B}}\ }%
  \textbf{\bibinfo {volume} {74}},\ \bibinfo {pages} {155402} (\bibinfo {year}
  {2006})%
  \bibAnnoteFile{NoStop}{chakarova2006}%
\bibitem{delle_site2003}%
  \BibitemOpen
  \bibfield{author}{%
  \bibinfo {author} {\bibfnamefont{L.}~\bibnamefont{Delle~Site}}, \bibinfo
  {author} {\bibfnamefont{A.}~\bibnamefont{Alavi}},\ and\ \bibinfo {author}
  {\bibfnamefont{C.~F.}\ \bibnamefont{Abrams}},\ }%
  \bibfield{journal}{%
  \Doi{10.1103/PhysRevB.67.193406}{\bibinfo {journal} {Physical Review B}}\ }%
  \textbf{\bibinfo {volume} {67}},\ \bibinfo {pages} {193406} (\bibinfo {year}
  {2003})%
  \bibAnnoteFile{NoStop}{delle_site2003}%
\bibitem{delle_site2004}%
  \BibitemOpen
  \bibfield{author}{%
  \bibinfo {author} {\bibfnamefont{L.}~\bibnamefont{Delle~Site}}, \bibinfo
  {author} {\bibfnamefont{S.}~\bibnamefont{Leon}},\ and\ \bibinfo {author}
  {\bibfnamefont{K.}~\bibnamefont{Kremer}},\ }%
  \bibfield{journal}{%
  \Doi{10.1021/ja0387406}{\bibinfo {journal} {Journal of the American Chemical
  Society}}\ }%
  \textbf{\bibinfo {volume} {126}},\ \bibinfo {pages} {2944} (\bibinfo {year}
  {2004})%
  \bibAnnoteFile{NoStop}{delle_site2004}%
\bibitem{delle_site2005}%
  \BibitemOpen
  \bibfield{author}{%
  \bibinfo {author} {\bibfnamefont{L.}~\bibnamefont{Delle~Site}}, \bibinfo
  {author} {\bibfnamefont{S.}~\bibnamefont{Leon}},\ and\ \bibinfo {author}
  {\bibfnamefont{K.}~\bibnamefont{Kremer}},\ }%
  \bibfield{journal}{%
  \Doi{10.1088/0953-8984/17/4/L01}{\bibinfo {journal} {Journal of Physics:
  Condensed Matter}}\ }%
  \textbf{\bibinfo {volume} {17}},\ \bibinfo {pages} {53} (\bibinfo {year}
  {2005})%
  \bibAnnoteFile{NoStop}{delle_site2005}%
\bibitem{johnston2007}%
  \BibitemOpen
  \bibfield{author}{%
  \bibinfo {author} {\bibfnamefont{K.}~\bibnamefont{Johnston}}\ and\ \bibinfo
  {author} {\bibfnamefont{R.~M.}\ \bibnamefont{Nieminen}},\ }%
  \bibfield{journal}{%
  \Doi{10.1103/PhysRevB.76.085402}{\bibinfo {journal} {Physical Review B}}\ }%
  \textbf{\bibinfo {volume} {76}},\ \bibinfo {pages} {085402} (\bibinfo {year}
  {2007})%
  \bibAnnoteFile{NoStop}{johnston2007}%
\bibitem{johnston2011}%
  \BibitemOpen
  \bibfield{author}{%
  \bibinfo {author} {\bibfnamefont{K.}~\bibnamefont{Johnston}}, \bibinfo
  {author} {\bibfnamefont{R.~M.}\ \bibnamefont{Nieminen}},\ and\ \bibinfo
  {author} {\bibfnamefont{K.}~\bibnamefont{Kremer}},\ }%
  \bibfield{journal}{%
  \Doi{10.1039/c1sm05410d}{\bibinfo {journal} {Soft Matter}}\ }%
  \textbf{\bibinfo {volume} {7}},\ \bibinfo {pages} {6457} (\bibinfo {year}
  {2011})%
  \bibAnnoteFile{NoStop}{johnston2011}%
\bibitem{moses2009}%
  \BibitemOpen
  \bibfield{author}{%
  \bibinfo {author} {\bibfnamefont{P.~G.}\ \bibnamefont{Moses}}, \bibinfo
  {author} {\bibfnamefont{J.~J.}\ \bibnamefont{Mortensen}}, \bibinfo {author}
  {\bibfnamefont{B.~I.}\ \bibnamefont{Lundqvist}},\ and\ \bibinfo {author}
  {\bibfnamefont{J.~K.}\ \bibnamefont{N{\o}rskov}},\ }%
  \bibfield{journal}{%
  \Doi{10.1063/1.3086040}{\bibinfo {journal} {The Journal of Chemical
  Physics}}\ }%
  \textbf{\bibinfo {volume} {130}},\ \bibinfo {pages} {104709} (\bibinfo {year}
  {2009})%
  \bibAnnoteFile{NoStop}{moses2009}%
\bibitem{wellendorff2010}%
  \BibitemOpen
  \bibfield{author}{%
  \bibinfo {author} {\bibfnamefont{J.}~\bibnamefont{Wellendorff}}
  \emph{et~al.},\ }%
  \bibfield{journal}{%
  \Doi{10.1007/s11244-010-9443-6}{\bibinfo {journal} {Topics in Catalysis}}\ }%
  \textbf{\bibinfo {volume} {53}},\ \bibinfo {pages} {378} (\bibinfo {year}
  {2010})%
  \bibAnnoteFile{NoStop}{wellendorff2010}%
\bibitem{bahn2002}%
  \BibitemOpen
  \bibfield{author}{%
  \bibinfo {author} {\bibfnamefont{S.~R.}\ \bibnamefont{Bahn}}\ and\ \bibinfo
  {author} {\bibfnamefont{K.~W.}\ \bibnamefont{Jacobsen}},\ }%
  \bibfield{booktitle}{%
  \emph{\bibinfo {booktitle} {Computing in Science \& Engineering}},\ }%
  \bibfield{journal}{%
  \Doi{10.1109/5992.998641}{\bibinfo {journal} {Computing in Science \&
  Engineering}}\ }%
  \textbf{\bibinfo {volume} {4}},\ \bibinfo {pages} {56} (\bibinfo {year}
  {2002})%
  \bibAnnoteFile{NoStop}{bahn2002}%
\bibitem{soler2002}%
  \BibitemOpen
  \bibfield{author}{%
  \bibinfo {author} {\bibfnamefont{J.~M.}\ \bibnamefont{Soler}} \emph{et~al.},\
  }%
  \bibfield{journal}{%
  \Doi{10.1088/0953-8984/14/11/302}{\bibinfo {journal} {Journal of Physics:
  Condensed Matter}}\ }%
  \textbf{\bibinfo {volume} {14}},\ \bibinfo {pages} {2745} (\bibinfo {year}
  {2002})%
  \bibAnnoteFile{NoStop}{soler2002}%
\bibitem{hammer1999}%
  \BibitemOpen
  \bibfield{author}{%
  \bibinfo {author} {\bibfnamefont{B.}~\bibnamefont{Hammer}}, \bibinfo {author}
  {\bibfnamefont{L.}~\bibnamefont{Hansen}},\ and\ \bibinfo {author}
  {\bibfnamefont{J.}~\bibnamefont{Nørskov}},\ }%
  \bibfield{journal}{%
  \Doi{10.1103/PhysRevB.59.7413}{\bibinfo {journal} {Physical Review B}}\ }%
  \textbf{\bibinfo {volume} {59}},\ \bibinfo {pages} {7413} (\bibinfo {year}
  {1999})%
  \bibAnnoteFile{NoStop}{hammer1999}%
\bibitem{perdew1996}%
  \BibitemOpen
  \bibfield{author}{%
  \bibinfo {author} {\bibfnamefont{J.~P.}\ \bibnamefont{Perdew}}, \bibinfo
  {author} {\bibfnamefont{K.}~\bibnamefont{Burke}},\ and\ \bibinfo {author}
  {\bibfnamefont{M.}~\bibnamefont{Ernzerhof}},\ }%
  \bibfield{journal}{%
  \Doi{10.1103/PhysRevLett.77.3865}{\bibinfo {journal} {Physical Review
  Letters}}\ }%
  \textbf{\bibinfo {volume} {77}},\ \bibinfo {pages} {3865} (\bibinfo {year}
  {1996})%
  \bibAnnoteFile{NoStop}{perdew1996}%
\bibitem{perdew1996a}%
  \BibitemOpen
  \bibfield{author}{%
  \bibinfo {author} {\bibfnamefont{J.~P.}\ \bibnamefont{Perdew}}, \bibinfo
  {author} {\bibfnamefont{K.}~\bibnamefont{Burke}},\ and\ \bibinfo {author}
  {\bibfnamefont{M.}~\bibnamefont{Ernzerhof}},\ }%
  \bibfield{journal}{%
  \Doi{10.1103/PhysRevLett.78.1396}{\bibinfo {journal} {Physical Review
  Letters}}\ }%
  \textbf{\bibinfo {volume} {78}},\ \bibinfo {pages} {1396} (\bibinfo {year}
  {1997})%
  \bibAnnoteFile{NoStop}{perdew1996a}%
\bibitem{blochl1994}%
  \BibitemOpen
  \bibfield{author}{%
  \bibinfo {author} {\bibfnamefont{P.~E.}\ \bibnamefont{Blöchl}},\ }%
  \bibfield{journal}{%
  \Doi{10.1103/PhysRevB.50.17953}{\bibinfo {journal} {Physical Review B}}\ }%
  \textbf{\bibinfo {volume} {50}},\ \bibinfo {pages} {17953} (\bibinfo {year}
  {1994})%
  \bibAnnoteFile{NoStop}{blochl1994}%
\bibitem{mortensen2005}%
  \BibitemOpen
  \bibfield{author}{%
  \bibinfo {author} {\bibfnamefont{J.~J.}\ \bibnamefont{Mortensen}}, \bibinfo
  {author} {\bibfnamefont{L.~B.}\ \bibnamefont{Hansen}},\ and\ \bibinfo
  {author} {\bibfnamefont{K.~W.}\ \bibnamefont{Jacobsen}},\ }%
  \bibfield{journal}{%
  \Doi{10.1103/PhysRevB.71.035109}{\bibinfo {journal} {Physical Review B}}\ }%
  \textbf{\bibinfo {volume} {71}},\ \bibinfo {pages} {035109} (\bibinfo {year}
  {2005})%
  \bibAnnoteFile{NoStop}{mortensen2005}%
\bibitem{enkovaara2010}%
  \BibitemOpen
  \bibfield{author}{%
  \bibinfo {author} {\bibfnamefont{J.}~\bibnamefont{Enkovaara}} \emph{et~al.},\
  }%
  \bibfield{journal}{%
  \Doi{10.1088/0953-8984/22/25/253202}{\bibinfo {journal} {Journal of Physics:
  Condensed Matter}}\ }%
  \textbf{\bibinfo {volume} {22}},\ \bibinfo {pages} {253202} (\bibinfo {year}
  {2010})%
  \bibAnnoteFile{NoStop}{enkovaara2010}%
\bibitem{monkhorst1976}%
  \BibitemOpen
  \bibfield{author}{%
  \bibinfo {author} {\bibfnamefont{H.~J.}\ \bibnamefont{Monkhorst}}\ and\
  \bibinfo {author} {\bibfnamefont{J.~D.}\ \bibnamefont{Pack}},\ }%
  \bibfield{journal}{%
  \Doi{10.1103/PhysRevB.13.5188}{\bibinfo {journal} {Physical Review B}}\ }%
  \textbf{\bibinfo {volume} {13}},\ \bibinfo {pages} {5188} (\bibinfo {year}
  {1976})%
  \bibAnnoteFile{NoStop}{monkhorst1976}%
\bibitem{pack1977}%
  \BibitemOpen
  \bibfield{author}{%
  \bibinfo {author} {\bibfnamefont{J.~D.}\ \bibnamefont{Pack}}\ and\ \bibinfo
  {author} {\bibfnamefont{H.~J.}\ \bibnamefont{Monkhorst}},\ }%
  \bibfield{journal}{%
  \Doi{10.1103/PhysRevB.16.1748}{\bibinfo {journal} {Physical Review B}}\ }%
  \textbf{\bibinfo {volume} {16}},\ \bibinfo {pages} {1748} (\bibinfo {year}
  {1977})%
  \bibAnnoteFile{NoStop}{pack1977}%
\bibitem{zhang1997}%
  \BibitemOpen
  \bibfield{author}{%
  \bibinfo {author} {\bibfnamefont{Y.}~\bibnamefont{Zhang}}\ and\ \bibinfo
  {author} {\bibfnamefont{W.}~\bibnamefont{Yang}},\ }%
  \bibfield{journal}{%
  \Doi{10.1103/PhysRevLett.80.890}{\bibinfo {journal} {Physical Review
  Letters}}\ }%
  \textbf{\bibinfo {volume} {80}},\ \bibinfo {pages} {890} (\bibinfo {year}
  {1998})%
  \bibAnnoteFile{NoStop}{zhang1997}%
\bibitem{roman-perez2009}%
  \BibitemOpen
  \bibfield{author}{%
  \bibinfo {author} {\bibfnamefont{G.~R.}\ \bibnamefont{P\'{e}rez}}\ and\
  \bibinfo {author} {\bibfnamefont{J.~M.}\ \bibnamefont{Soler}},\ }%
  \bibfield{journal}{%
  \Doi{10.1103/PhysRevLett.103.096102}{\bibinfo {journal} {Physical Review
  Letters}}\ }%
  \textbf{\bibinfo {volume} {103}},\ \bibinfo {pages} {096102} (\bibinfo {year}
  {2009})%
  \bibAnnoteFile{NoStop}{roman-perez2009}%
\bibitem{dion2004}%
  \BibitemOpen
  \bibfield{author}{%
  \bibinfo {author} {\bibfnamefont{M.}~\bibnamefont{Dion}}, \bibinfo {author}
  {\bibfnamefont{H.}~\bibnamefont{Rydberg}}, \bibinfo {author}
  {\bibfnamefont{E.}~\bibnamefont{Schr\"{o}der}}, \bibinfo {author}
  {\bibfnamefont{D.~C.}\ \bibnamefont{Langreth}},\ and\ \bibinfo {author}
  {\bibfnamefont{B.~I.}\ \bibnamefont{Lundqvist}},\ }%
  \bibfield{journal}{%
  \Doi{10.1103/PhysRevLett.92.246401}{\bibinfo {journal} {Physical Review
  Letters}}\ }%
  \textbf{\bibinfo {volume} {92}},\ \bibinfo {pages} {246401} (\bibinfo {year}
  {2004})%
  \bibAnnoteFile{NoStop}{dion2004}%
\bibitem{dion2005}%
  \BibitemOpen
  \bibfield{author}{%
  \bibinfo {author} {\bibfnamefont{M.}~\bibnamefont{Dion}}, \bibinfo {author}
  {\bibfnamefont{H.}~\bibnamefont{Rydberg}}, \bibinfo {author}
  {\bibfnamefont{E.}~\bibnamefont{Schröder}}, \bibinfo {author}
  {\bibfnamefont{D.~C.}\ \bibnamefont{Langreth}},\ and\ \bibinfo {author}
  {\bibfnamefont{B.~I.}\ \bibnamefont{Lundqvist}},\ }%
  \bibfield{journal}{%
  \Doi{10.1103/PhysRevLett.95.109902}{\bibinfo {journal} {Physical Review
  Letters}}\ }%
  \textbf{\bibinfo {volume} {95}},\ \bibinfo {pages} {109902} (\bibinfo {year}
  {2005})%
  \bibAnnoteFile{NoStop}{dion2005}%
\bibitem{guo1992}%
  \BibitemOpen
  \bibfield{author}{%
  \bibinfo {author} {\bibfnamefont{J.}~\bibnamefont{Guo}}, \bibinfo {author}
  {\bibfnamefont{D.~E.}\ \bibnamefont{Ellis}},\ and\ \bibinfo {author}
  {\bibfnamefont{D.~J.}\ \bibnamefont{Lam}},\ }%
  \bibfield{journal}{%
  \Doi{10.1103/PhysRevB.45.13647}{\bibinfo {journal} {Physical Review B}}\ }%
  \textbf{\bibinfo {volume} {45}},\ \bibinfo {pages} {13647} (\bibinfo {year}
  {1992})%
  \bibAnnoteFile{NoStop}{guo1992}%
\bibitem{toofan1998}%
  \BibitemOpen
  \bibfield{author}{%
  \bibinfo {author} {\bibfnamefont{J.}~\bibnamefont{Toofan}}\ and\ \bibinfo
  {author} {\bibfnamefont{P.~R.}\ \bibnamefont{Watson}},\ }%
  \bibfield{journal}{%
  \Doi{10.1016/S0039-6028(97)01031-5}{\bibinfo {journal} {Surface Science}}\ }%
  \textbf{\bibinfo {volume} {401}},\ \bibinfo {pages} {162} (\bibinfo {year}
  {1998})%
  \bibAnnoteFile{NoStop}{toofan1998}%
\end{thebibliography}%

\end{document}